\documentclass[10pt,conference]{IEEEtran}
\usepackage{amsmath}
\newtheorem{theorem}{Theorem}[section]

\newtheorem{definition}[theorem]{Definition}
\newtheorem{lemma}[theorem]{Lemma}
\newtheorem{example}[theorem]{Example}
\newtheorem{corollary}[theorem]{Corollary}
\newtheorem{remark}[theorem]{Remark}

\font\msbm=msbm10 at 10pt

\newcommand{\FF}{\mbox{\msbm F}}
\newcommand{\RR}{\mbox{\msbm R}}

\newcommand{\CC}{\mbox{\msbm C}}

\begin{document}

\title{Matrix Codes as Ideals for Grassmannian Codes and their Weight Properties}
\author{
   \IEEEauthorblockN{Bryan S. Hernandez and Virgilio P. Sison}
   \IEEEauthorblockA{Institute of Mathematical Sciences and Physics\\
     University of the Philippines, Los Ba\~{n}os\\
     College, Laguna 4031, Philippines\\
     Email: \{bshernandez, vpsison\}@up.edu.ph} 
 }
\maketitle
\begin{abstract}
\boldmath A systematic way of constructing Grassmannian codes endowed with the subspace distance as lifts of matrix codes over the prime field $\FF_p$ is introduced. The matrix codes are $\FF_p$-subspaces of the ring $M_2(\FF_p)$ of $2 \times 2$ matrices over $\FF_p$ on which the rank metric is applied, and are generated as one-sided proper principal ideals by idempotent elements of $M_2(\FF_p)$. Furthermore a weight function on the non-commutative matrix ring $M_2(\FF_q)$, $q$ a power of $p$, is studied in terms of the egalitarian and homogeneous conditions. The rank weight distribution of $M_2(\FF_q)$ is completely determined by the general linear group $GL(2,q)$. Finally a weight function on subspace codes is analogously defined and its egalitarian property is examined.
\end{abstract}
\vskip .1in

\section{Introduction}
\label{sec:int}
Certain concepts of ``coding theory in projective space" and the practical significance of subspace codes in error correction in networks are highlighted in this paper. Let $q=p^r$, $p$ a prime, $r$ a positive integer, and $\FF_q$ the Galois field with cardinality $q$ and characteristic $p$. Consider the $n$-dimensional full vector space $\FF_q^n$ over $\FF_q$. The set of all subspaces of $\FF_q^n$, denoted by $\mathcal P_q(n)$, is called the projective space of order $n$ over $\FF_q$. For an integer $k$, where $0 \le k \le n$, the set of all $k$-dimensional subspaces of $\FF_q^n$, denoted by $\mathcal G_q(n,k)$, is called the Grassmannian. A subspace code is a nonempty subset of $\mathcal P_q(n)$. A Grassmannian code is a nonempty subset of $\mathcal G_q(n,k)$ which is also called a constant dimension code, that is, the codewords in $\mathcal G_q(n,k)$ are subspaces of $\FF_q^n$ of dimension $k$, thus they are nothing but rate-$k/n$ linear block codes of length $n$ over $\FF_q$. Subspace codes have practical importance in network coding. The seminal paper \cite{ahlswede} refers to \textit{network coding} as ``coding at a node in a network", that is, a node receives information from all input links, then encodes and sends information to all output links.

This present work deals mainly with the linear construction of Grassmannian codes endowed with the subspace distance from lifts of matrix codes over $\FF_p$ which are seen as one-sided principal ideals generated by the idempotent elements of the non-commutative matrix ring $M_2(\FF_p)$. The matrix codes are endowed with the so-called rank weight, which is not egalitarian nor homogeneous, but nevertheless is completely determined by the multiplicative group of invertible matrices.  

The second section of this paper gives the theoretical requisites. Examples of rank-metric codes and Grassmannian codes from left (resp. right) ideals of $M_2(\FF_p)$ using idempotent elements of $M_2(\FF_p)$ are given in Section \ref{grass:ideals}. Section \ref{rank:weight} discusses the weight properties of rank metric codes, while Section \ref{subs:prop} studies the weight properties of the associated subspace codes. 

\section{Preliminaries}
\label{prelim}

\begin{definition}
Let $R$ be a ring and $\RR$ be the set of real numbers. A mapping $w:R  \rightarrow \RR$ is called a {\it weight} if the following conditions are satisfied:
	\begin{enumerate}
    	\item[i.] $w(x)=0$       if and only if  $x = 0$, for all $x \in R$;
    	\item[ii.] $w(x) \geq 0$      for all $x \in R$; 
    	\item[iii.] $w(x)=w(-x)$,      for all $x \in R$; and
	\item[iv.] $w(x+y) \leq w(x) + w(y)$,      for all $x,y \in R$.
    	\end{enumerate}
\end{definition}

\begin{definition}
A weight $w$ on the finite ring $R$ is said to be {\it egalitarian} if satisfies condition (E) as follows.
\begin{enumerate}
\item[(E)] there exists a constant $\Gamma$ such that \[\sum\limits_{y \in Rx} {w(y)}  = \Gamma \left| {Rx} \right|\] for all $x \in R\backslash\{0\}$.
\end{enumerate}
The weight $w$ is said to be (left) {\it homogeneous} if it satisifies (E) and the additional condition (H) as follows. 
\begin{enumerate}
\item[(H)] $w(x)=w(y)$ for all $x,y \in R$ such that $Rx=Ry$.
\end{enumerate}
\end{definition}

The definition for a right homogeneous weight follows analogously.  If $w$ is both left and right homogeneous, it is said to be {\it homogeneous}. The number $\Gamma$ is called the {\it average value} of $w$. The weight $w$ is said to be {\it normalized} if $\Gamma=1$.

The set of all $k\times\ell$ matrices over $\FF_q$, denoted by $M_{k\times\ell}(\FF_q)$, is considered as a vector space over $\FF_q$. A nonempty subset of $M_{k \times \ell}(\FF_q)$ is called a $[k \times \ell]$ {\it matrix code} over $\FF_q$. This $[k\times\ell]$ matrix code is said to be linear if it is a subspace of $M_{k \times \ell}(\FF_q)$.

The {\it rank distance} between two $k \times \ell$ matrices over $\FF_q$, say $A$ and $B$, is defined by $d_R(A,B)= $ rank$(A-B)$, and is clearly a metric. A $[k \times \ell,\delta]$ {\it rank-metric code} $\CC$ is a $[k \times \ell]$ matrix code whose minimum rank distance is $\delta$. That is, $$\delta=\min\{d_R(A,B)|A,B \in \CC, A\neq B\}.$$ 

\begin{definition}
A $[k\times \ell,\rho,\delta]$ {\it rank-metric code} is a linear code in $M_{k \times \ell} (\FF_q)$ with dimension $\rho$ and minimum rank distance $\delta$. 
\end{definition}

\begin{definition}
Let $A \in M_{k \times \ell} (\FF_q)$. The {\it{lift}} of $A$, denoted by $L(A)$, is the $k\times(k+\ell)$ standard matrix $(I_k$ $A)$, where $I_k$ is the $k \times k$ identity matrix.
\end{definition}

The subspace generated by the rows of the lifted matrix $L(A)$ will be denoted by $\langle L(A) \rangle$. This subspace is in fact a rate-$k/(k+\ell)$ linear block code of length $k + \ell$ over $\FF_q$.

The matrix ring of all $2\times 2$ matrices over $\FF_p$, denoted by $M_2(\FF_p)$, has no proper two sided ideals but it has proper left sided ideals \cite{hun}. It has $p+1$ minimal left ideals and each minimal left ideal contains $p^2$ elements. These minimal left ideals are themselves the maximal left ideals \cite{falcunit:sison}. The left ideals are easily seen as linear codes in $M_2(\FF_p)$. Certainly we get similar results for the minimal right ideals. The theorem below shows that the proper left ideals are generated by the idempotent elements of $M_2(\FF_p)$.

\begin{theorem}[Falcunit and Sison, \cite{falcunit:sison}]
Each minimal left ideal of $M_2(\FF_{p})$ takes the form $$M_2(\FF_{p})a=\{ra|r \in M_2(\FF_{p})\},$$ where $a$ is a nonzero nonunit idempotent of $M_2(\FF_{p})$, and it contains $p^2$ elements.
\end{theorem}

There are $p+1$ nonzero nonunit idempotents of $M_2(\FF_{p})$. These are $ {\begin{pmatrix}
                                                      0 & 0  \\
                                                      0 & 1 
                                           \end{pmatrix}}$ and $ {\begin{pmatrix}
                                                      1 & r  \\
                                                      0 & 0 
                                           \end{pmatrix}}$ where $r \in \FF_{p}$ \cite{falcunit:sison}.

We consider the rank distance for the linear code $M_2(\FF_{p})a=\{ra|r \in M_2(\FF_{p})\}$ where $a$ is a nonzero nonunit idempotent of $M_2(\FF_{p})$.

On the projective space $\mathcal P_q(n)$ there are at least two metrics that can be applied. The {\it{subspace distance}}  is given by $$ d_S(A,B)=\dim A + \dim B - 2\dim(A \cap B)$$ while the next one is the {\it{injection distance}} given by $$d_I(A,B)=\max\{\dim A ,\dim B\} - \dim(A \cap B),$$ for $A,B \in \mathcal P_q(n)$. In this paper we shall only use the subspace distance on the constructed Grassmannian codes.

A classic formula for the cardinality of the Grassmannian $\mathcal G _q(n,k)$ is given by the $q$-ary Gaussian coefficient $$ {n \brack k}_q = \prod^{k-1}_{i=0} \frac{q^n-q^i}{q^k-q^i}.$$ 

\begin{definition}
A subset $\mathcal C$ of $\mathcal G _q (n,k)$ is an $(n,M,d,k)_q$  {\it{code in the Grassmannian}} if $|\mathcal C|=M$ and \[d=\{\min d_S(U,V)|U,V \in \mathcal C, U\neq V\}.\] 
\end{definition}

\begin{definition}
Let $\mathcal C$ be a $[k \times \ell]$ rank-metric code. The set 
\begin{align*}
\Lambda (\mathcal C)=\{\langle L(A)  \rangle|A \in \mathcal C \}
\end{align*}
is called the {\it{lift}} of $\mathcal C$.
\end{definition}

\begin{theorem}{(T. Etzion, \cite{etz})}
\label{rank:metric}
Let $\mathcal C$ be a $[k \times \ell,\rho,\delta]$ rank-metric code. The lift of  $\mathcal C$ is a $(k+\ell, q^\rho,2\delta, k)_q$ Grassmannian code.
\end{theorem}

\section{Rank-Metric Codes and Grassmannian Codes from One Sided Ideals of $M_2(\FF_p)$}
\label{grass:ideals}

In this section we construct new examples of Grassmannian codes from rank-metric codes which are left or right ideals generated by  idempotent elements of $M_2(\FF_p)$. An idempotent element of $M_2(\FF_p)$ is carefully chosen to yield a left or right ideal to obtain a rank-metric code which is subsequently lifted to form a Grassmannian code. The parameters of the associated Grassmannian code are given in the theorem below.

\begin{theorem}
\label{idemp}
Let $a$ is a nonzero nonunit idempotent of $M_2(\FF_{p})$ and $M_2(\FF_{p})a=\{ra|r \in M_2(\FF_{p})\}$.
Then $\Lambda(M_2(\FF_{p}))$ is a $(4,p^2,2,2)_p$ Grassmannian code.
\end{theorem}

Correspondingly, we can consider the right ideal $aM_2(\FF_{p})=\{ar|r \in M_2(\FF_{p})\}$ for a nonzero nonunit idempotent $a \in M_2(\FF_{p})$.

\begin{example}
\label{left:ideal1}
Consider an idempotent element $ {\begin{pmatrix}
                                                      0 & 0  \\
                                                      0 & 1 
                                           \end{pmatrix}} \in M_2(\FF_2)$. We generate the left ideal $I$ using the given idempotent element.
\[\begin{aligned}
 I&=\left \{ \left.
r
{\begin{pmatrix}
                                                      0 & 0  \\
                                                      0 & 1 
                                           \end{pmatrix}}
\right| r \in M_2(\FF_2) \right \}\\ 
&=\left \{ \left.
 {\begin{pmatrix}
                                                      a_0 & a_1  \\
                                                      a_2 & a_3 
                                           \end{pmatrix}}
{\begin{pmatrix}
                                                      0 & 0  \\
                                                      0 & 1 
                                           \end{pmatrix}}
\right| a_i \in \FF_2 \right \}\\
&=\left \{ \left.
 {\begin{pmatrix}
                                                      0 & a_1  \\
                                                      0 & a_3 
                                           \end{pmatrix}}
\right| a_i \in \FF_2 \right \}\\
&=\left \{ 
 {\begin{pmatrix}
                                                      0 & 0  \\
                                                      0 & 0 
                                           \end{pmatrix}},
 {\begin{pmatrix}
                                                      0 & 1  \\
                                                      0 & 0 
                                           \end{pmatrix}},
 {\begin{pmatrix}
                                                      0 & 0  \\
                                                      0 & 1 
                                           \end{pmatrix}},
 {\begin{pmatrix}
                                                      0 & 1  \\
                                                      0 & 1 
                                           \end{pmatrix}}
\right \}.
\end{aligned}\]
Note that $I$ is a $[2 \times 2, 2, 1]$ rank-metric code with values $k=2, \ell = 2, \rho = 2$, and $\delta =1$. Now the lifted matrices are 
\[
 {\begin{pmatrix}
                                                   1 &  0 & 0 & 0  \\
                                                   0 &  1 &   0 & 0 
                                           \end{pmatrix}},
 {\begin{pmatrix}
                                                   1 &  0 &   0 & 1  \\
                                                    0 &  1 &  0 & 0 
                                           \end{pmatrix}},
 {\begin{pmatrix}
                                                    1 &  0 &  0 & 0  \\
                                                    0 &  1 &  0 & 1 
                                           \end{pmatrix}},
  \]
 ${\text {and } }  {\begin{pmatrix}
                                                   1 &  0 &   0 & 1  \\
                                                   0 &  1 &   0 & 1 
                                           \end{pmatrix}}.$
We then get the following subspaces generated by the rows of the lifted matrices.
 \[\begin{aligned}
C_1 & =\{(1,0,0,0),(0,1,0,0),(1,1,0,0),(0,0,0,0)\};\\
C_2 & =\{(1,0,0,1),(0,1,0,0),(1,1,0,1),(0,0,0,0)\};\\
C_3 & =\{(1,0,0,0),(0,1,0,1),(1,1,0,1),(0,0,0,0)\}; \\
C_4 & =\{(1,0,0,1),(0,1,0,1),(1,1,0,0),(0,0,0,0)\}.
\end{aligned}\]
The lifted rank-metric code given by $ \{ C_1,C_2,C_3,C_4 \} $ is a $(4,4,2,2)_2$ Grassmannian code by Theorems \ref{rank:metric} and \ref{idemp}.
\end{example}

\begin{example}
\label{right:ideal1}
Consider the same idempotent element $ {\begin{pmatrix}
                                                      0 & 0  \\
                                                      0 & 1 
                                           \end{pmatrix}} \in M_2(\FF_2)$. This time we generate the right ideal $\overline I$ using the given idempotent element.
\[\begin{aligned}
 \overline I &=\left \{ \left.
{\begin{pmatrix}
                                                      0 & 0  \\
                                                      0 & 1 
                                           \end{pmatrix}}r
\right| r \in M_2(\FF_2) \right \}\\ 
&=\left \{ \left.
{\begin{pmatrix}
                                                      0 & 0  \\
                                                      0 & 1 
                                           \end{pmatrix}}
 {\begin{pmatrix}
                                                      a_0 & a_1  \\
                                                      a_2 & a_3 
                                           \end{pmatrix}}
\right| a_i \in \FF_2 \right \}\\
&=\left \{ \left.
 {\begin{pmatrix}
                                                      0 & 0  \\
                                                      a_2 & a_3 
                                           \end{pmatrix}}
\right| a_i \in \FF_2 \right \}\\
&=\left \{ 
 {\begin{pmatrix}
                                                      0 & 0  \\
                                                      0 & 0 
                                           \end{pmatrix}},
 {\begin{pmatrix}
                                                      0 & 0  \\
                                                      1 & 0 
                                           \end{pmatrix}},
 {\begin{pmatrix}
                                                      0 & 0  \\
                                                      0 & 1 
                                           \end{pmatrix}},
 {\begin{pmatrix}
                                                      0 & 0  \\
                                                      1 & 1 
                                           \end{pmatrix}}
\right \}.
\end{aligned}\]
Note that $\overline I$ is a $[2 \times 2, 2, 1]$ rank-metric code with values $k=2, \ell = 2, \rho = 2$, and $\delta =1$. The lifted matrices are 
\[
 {\begin{pmatrix}
                                                   1 &  0 & 0 & 0  \\
                                                   0 &  1 &   0 & 0 
                                           \end{pmatrix}},
 {\begin{pmatrix}
                                                   1 &  0 &   0 & 0  \\
                                                    0 &  1 &  1 & 0 
                                           \end{pmatrix}},
 {\begin{pmatrix}
                                                    1 &  0 &  0 & 0  \\
                                                    0 &  1 &  0 & 1 
                                           \end{pmatrix}},
\]
${\text {and } }  {\begin{pmatrix}
                                                   1 &  0 &   0 & 0  \\
                                                   0 &  1 &   1 & 1 
                                           \end{pmatrix}}.$
The subspaces generated by the rows of the lifted matrices are given by
 \[\begin{aligned}
C_1 & =\{(1,0,0,0),(0,1,0,0),(1,1,0,0),(0,0,0,0)\};\\
C_2 & =\{(1,0,0,0),(0,1,1,0),(1,1,1,0),(0,0,0,0)\};\\
C_3 & =\{(1,0,0,0),(0,1,0,1),(1,1,0,1),(0,0,0,0)\}; \\
C_4 & =\{(1,0,0,0),(0,1,1,1),(1,1,1,1),(0,0,0,0)\}.
\end{aligned}\]
Hence the lifted code is given by $ \{ C_1,C_2,C_3,C_4 \}$ which is a $(4,4,2,2)_2$ Grassmannian code.
\end{example}

Note that the left and right ideals generated by the same idempotent element in Example \ref{left:ideal1} and Example \ref{right:ideal1}, respectively, are different yet the lifts of the rank-metric codes are both $(4,4,2,2)_2$ Grassmannian codes.

\begin{example}
Given an idempotent element of $M_2(\FF_3)$: $ {\begin{pmatrix}
                                                      0 & 2  \\
                                                      0 & 1 
                                           \end{pmatrix}}$. The left ideal generated by this idempotent element is
\[\begin{aligned}
 J&=\left \{ \left.
r
{\begin{pmatrix}
                                                      0 & 2  \\
                                                      0 & 1 
                                           \end{pmatrix}}
\right| r \in M_2(\FF_3) \right \}\\ 
&=\left \{ \left.
 {\begin{pmatrix}
                                                      a_0 & a_1  \\
                                                      a_2 & a_3 
                                           \end{pmatrix}}
{\begin{pmatrix}
                                                      0 & 2  \\
                                                      0 & 1 
                                           \end{pmatrix}}
\right| a_i \in \FF_3 \right \}\\
&=\left \{ \left.
{\begin{pmatrix}
                                                      0 & 2a_0+a_1  \\
                                                      0 & 2a_2+a_3 
                                           \end{pmatrix}}
\right| a_i \in \FF_3 \right \}.
\end{aligned}\]
Hence we have $ J=\left \{ X_1, X_2, X_3, X_4, X_5, X_6, X_7, X_8, X_9 \right \} $ where 
\[\begin{aligned}
 X_1={\begin{pmatrix}
                                                      0 & 0  \\
                                                      0 & 0 
                                           \end{pmatrix}},
  X_2={\begin{pmatrix}
                                                      0 & 1  \\
                                                      0 & 0 
                                           \end{pmatrix}},
 X_3={\begin{pmatrix}
                                                      0 & 2  \\
                                                      0 & 0 
                                           \end{pmatrix}},\\
  X_4={\begin{pmatrix}
                                                      0 & 0  \\
                                                      0 & 1 
                                           \end{pmatrix}},
 X_5= {\begin{pmatrix}
                                                      0 & 0  \\
                                                      0 & 2 
                                           \end{pmatrix}},
X_6={\begin{pmatrix}
                                                      0 & 1  \\
                                                      0 & 1 
                                           \end{pmatrix}},\\
 X_7= {\begin{pmatrix}
                                                      0 & 2  \\
                                                      0 & 2 
                                           \end{pmatrix}},
 X_8= {\begin{pmatrix}
                                                      0 & 2  \\
                                                      0 & 1 
                                           \end{pmatrix}},
 X_9= {\begin{pmatrix}
                                                      0 & 1  \\
                                                      0 & 2 
                                           \end{pmatrix}}.
\end{aligned}\]
Note that $J$ is a $[2 \times 2, 2, 1]$ rank-metric code with values $k=2, \ell = 2, \rho = 2$, and $\delta =1$. The lifted matrices are 
\[ 
 {\begin{pmatrix}
                                                   1 &  0 & 0 & 0  \\
                                                   0 &  1 &   0 & 0 
                                           \end{pmatrix}},
 {\begin{pmatrix}
                                                   1 &  0 &   0 & 1  \\
                                                    0 &  1 &  0 & 0 
                                           \end{pmatrix}},
 {\begin{pmatrix}
                                                   1 &  0 & 0 & 2  \\
                                                   0 &  1 &   0 & 0 
                                           \end{pmatrix}},\]
\[{\begin{pmatrix}
                                                   1 &  0 &   0 & 0  \\
                                                    0 &  1 &  0 & 1 
                                           \end{pmatrix}},
 {\begin{pmatrix}
                                                    1 &  0 &  0 & 0  \\
                                                    0 &  1 &  0 & 2 
                                           \end{pmatrix}},
 {\begin{pmatrix}
                                                   1 &  0 & 0 & 1  \\
                                                   0 &  1 &   0 & 1 
                                           \end{pmatrix}},\]
 \[{\begin{pmatrix}
                                                   1 &  0 &   0 & 2  \\
                                                    0 &  1 &  0 & 2 
                                           \end{pmatrix}},
 {\begin{pmatrix}
                                                    1 &  0 &  0 & 2  \\
                                                    0 &  1 &  0 & 1 
                                           \end{pmatrix}},
 {\begin{pmatrix}
                                                   1 &  0 &   0 & 1  \\
                                                   0 &  1 &   0 & 2 
                                           \end{pmatrix}}.\]
By Theorems \ref{rank:metric} and \ref{idemp}, the lifted rank-metric code is a $(4,9,2,2)_2$ Grassmannian code.
\end{example}

\section{Rank-Metric Codes and their Weight Properties}
\label{rank:weight}
Consider the non-commutative ring $M_n(\FF_{q})$ of $n \times n$ matrices over $\FF_q$. The rank of $A \in M_n(\FF_{q})$ can be seen as a weight function from $M_n(\FF_{q})$ to $\RR$. We shall call this the rank weight of $A$.
\begin{theorem}
Let $A \in M_n(\FF_{q})$. The function $w_R$ from $M_n(\FF_{q})$ to $\RR$, defined by $w_R(A)= {\text{rank}}(A)$, is a weight. \end{theorem}

Let $\mathcal C$ be a $[k \times \ell, \rho, \delta]$ rank-metric code. The minimum rank weight of $\mathcal C$, denoted by $\Omega$, is the smallest nonzero rank among its elements, that is, $\Omega=\min \{w_R(A)|A \in  \mathcal C, A \neq 0\}.$
\begin{theorem}
Let $\mathcal C$ be a $[k \times \ell, \rho, \delta]$ rank-metric code and $\Omega$ be its minimum rank weight. Then $\delta=\Omega$.
\label{delta:omega}
\end{theorem}
\begin{IEEEproof} 
Let $\mathcal C$ be a rank-metric code with minimum rank distance $\delta$ and minimum rank weight $\Omega$. Let $A$ and $B$ be distinct elements of $\mathcal C$ such that ${\text{rank}}(A-B)$ is minimum. Note that ${\text{rank}}(A-B) \ne 0$.  Then $\delta=d_R(A,B)=w_R(A-B) \ge \Omega$. Moreover, let $A \in \mathcal C$ with minimum rank. Now, $\Omega=w_R(A)=d_R(A,0) \ge \delta$. Thus, $\delta = \Omega$.
\end{IEEEproof} 

\begin{example}
Consider the following rank-metric code given in Example \ref{left:ideal1}:
\[
I=\left \{ 
 {\begin{pmatrix}
                                                      0 & 0  \\
                                                      0 & 0 
                                           \end{pmatrix}},
 {\begin{pmatrix}
                                                      0 & 1  \\
                                                      0 & 0 
                                           \end{pmatrix}},
 {\begin{pmatrix}
                                                      0 & 0  \\
                                                      0 & 1 
                                           \end{pmatrix}},
 {\begin{pmatrix}
                                                      0 & 1  \\
                                                      0 & 1 
                                           \end{pmatrix}}
\right \}.\]
Note that the minimum rank weight of $I$ is 1. By Theorem \ref{delta:omega}, the minimum rank distance of $I$ is also 1.
\end{example}

\begin{lemma}
The rank weight on $M_2(\FF_{q})$ has the explicit form below.  
\begin{gather*}
		   {\tt{rank}}(A) = \begin{cases}
       		 0 & {\rm if} \ A = {0} \\
     		 1& {\rm if} \ A {\text { is a zero divisor. } } \\
      		 2 &  {\rm if} \ A {\text { is a unit } } \\
      		 \end{cases}
	\end{gather*}
\label{partition}
for all $A \in M_2(\FF_{q}).$
\end{lemma}

Let $A_i$ be the number of elements of $M_2(\FF_q)$ with rank $i$ where $0 \le i \le n$. Consequently, $M_2(\FF_q)$ has the following rank distribution.
\begin{itemize}
\item[i.] $A_0=1$;
\item[ii.] $A_1=q^4-|GL(2,q)|-1$; and,
\item[iii.] $A_2=|GL(2,q)|$.
\end{itemize}

\begin{theorem}
The rank weight $w_R$ on $M_2(\FF_{p})$ is not egalitarian nor homogeneous.
\end{theorem}
\begin{IEEEproof} 

Let $\mathcal R=M_2(\FF_{p})=\left \{ 
 {\begin{pmatrix}
                                                      a_1 & a_3  \\
                                                      a_2 & a_4 
                                           \end{pmatrix}}
  \Big | a_i \in \FF_p \ \right \}$ .

Consider $y= 
 {\begin{pmatrix}
                                                      0 & 0  \\
                                                      0 & 1 
                                           \end{pmatrix}}$,
an idempotent element of $\mathcal R$ to form the minimal left ideal  $ \mathcal R y=\left \{ 
 {\begin{pmatrix}
                                                      0 & a_1  \\
                                                      0 & a_2 
                                           \end{pmatrix}}
\Big | a_1,a_3 \in \FF_p \right \}$ of $\mathcal R $.  For $w_R$ to be egalitarian, there must exist a unique $\Gamma \in \RR$ such that  $\Gamma= \dfrac{ \displaystyle \sum_{A\in \mathcal R x}{w_r(A)}}{| \mathcal R x|}$ for any left ideal $ \mathcal R x$ of $\mathcal R $.
\vskip.1in
We have $\sum_{A\in M_2(\FF_p)} w_R(A) = |GL(2,p)|(2)+(p^4-|GL(2,p)|-1)(1)+1(0)=2p^4-p^3-p^2+p-1.$

Let $\Gamma_1=\dfrac{ \displaystyle \sum_{A\in \mathcal R}{w_r(A)}}{| \mathcal R |}$. Then
$\Gamma_1=\dfrac{2p^4-p^3-p^2+p-1}{p^4}$.

Also we have
$\displaystyle \sum_{A\in  \mathcal R y}{w_r(A)}
=p^2-1.$ Now let 

$\Gamma_2=\dfrac{ \displaystyle \sum_{A\in  \mathcal Ry}{w_r(A)}}{| \mathcal R y |}$. Then
$\Gamma_2=\dfrac{p^2-1}{p^2}$. Note that 
\[\begin{aligned}
\dfrac{p^2-1}{p^2}
&=\dfrac{2p^4-p^3-p^2+p-1+(-p^4+p^3-p+1)}{p^4}.
\end{aligned}\]

But $\Gamma_1$ will only be equal to $\Gamma_2$ if and only if \[p^4-p^3+p-1=(p^3+1)(p-1)=0.\] Clearly there does not exist a prime $p$ that satisfies the obtained equation. Thus the rank weight $w_R$ on $M_2(\FF_p)$ is not egalitarian and it cannot be homogeneous as well. \end{IEEEproof} 

\begin{remark}
In general, for a positive integer $n$, the rank weight $w_R$ on $M_n(\FF_{q})$ is not homogeneous. Further the rank weight $w_R$ is non-egalitarian yet it satisfies the following property.
\end{remark}

\begin{lemma}
Let $A \in M_2(\FF_{q})$ and $U \in GL(2,q)$. Then $w_R(A)=w_R(UA)$.
\end{lemma}

\section{Subspace Codes and their Weight Properties}
\label{subs:prop}
The dimension of $A \in \mathcal P_q(n)$ can be seen as a weight function $w_S$ from $\mathcal P_q(n)$ to $\RR$. We shall call this the subspace weight of $A$. The fact that, $-A=\{-a|a \in A\}=A$, since $A$ is an additive group, will be useful in the following theorem.  

\begin{theorem}
The function $w_S$ from $\mathcal P_q(n)$ to $\RR$, defined by $w_S(A)= \dim (A),$ is a weight.
\label{dimweight}
\end{theorem}

\begin{definition}
Let $\mathcal C$ be a subspace code in $P_q(n)$. The minimum subspace weight of $\mathcal C$, denoted by $\Delta$, is the smallest nonzero dimension among its elements, that is, $$\Delta=\min \{w_S(A)|A \in  \mathcal C {\text{ and }} A \neq 0\}.$$
\end{definition}

\begin{remark}
If $\mathcal C$ is a subspace code in $P_q(n)$ of minimum subspace weight $\Delta$, then clearly $1 \le \Delta \le n$.
\end{remark}

\begin{theorem}
\label{subs}
Consider the projective space $\mathcal P_q(n)$ of order $n$ over $\FF_q$. Then $A+B \in \mathcal P_q(n)$ for all $A,B \in \mathcal P_q(n)$. \end{theorem}
\begin{IEEEproof} We have $A+B = \{(a_1,a_2,...,a_n)+(b_1,b_2,...,b_n)|a_i \in A, b_i \in B\}$. Let $A,B \in \mathcal P_q(n)$. Since $A$ and $B$ are subspaces of $\FF_q^n$, the zero vector is an element of both $A$ and $B$ and hence $A+B \ne \emptyset$. Let $x,y \in A+B$ and $r \in \FF_q$. Now, $x=a+b$ and $y=c+d$ for some $a,c \in A$ and $b,d \in B$. We have,
\[\begin{aligned}
x+ry
&=a+b + r(c+d) \\ &=(a+rc)+(b+rd) \in A+B. 
\end{aligned}\] 
Thus, $A+B$ is a subspace of $\FF_q^n$. \end{IEEEproof}

\begin{example}
\label{subs:weig}
Consider the subspace code $\mathcal C$ in $\mathcal P_2(3)$ given by 
$\mathcal C=\{A,B,A+B\}$
where \[A=\{(0,0,0),(1,0,1),(0,1,0),(1,1,1)\},\]
\[B=\{(0,0,0),(0,1,1),(1,0,0),(1,1,1)\} \] 
\[\begin{aligned}
A+B=&\{(0,0,0),(1,0,1),(0,1,0),(1,1,1),\\
&(0,1,1),(1,0,0),(0,0,1),(1,1,0)\}.
\end{aligned}\]
Note that $A+B$ is the entire space $\FF_2^3$. The minimum weight of $\mathcal C$ is $\dim(A)=\dim(B)=2$ and its minimum distance is $$\dim(A+B)-\dim(A)-2\dim((A+B) \cap A)=3+2-2(2)=1.$$ In this case, $d$ is not equal to $\Delta$.
\end{example}

Example \ref{subs:weig} shows that given a subspace code $\mathcal C$ such that $A+B \in \mathcal C$ for all $A, B \in \mathcal C$, the minimum subspace distance $d$ is not equal to the minimum weight $\Delta$. This means that even if we impose the condition $A+B \in \mathcal C$ for all $A, B \in \mathcal C$, $d \ne \Delta$ nor $d \ge \Delta$.

The following theorem gives under certain conditions a formula that computes the minimum distance of a subspace code in terms of the subspace weight. 

\begin{theorem}
Let $\mathcal C$ be a subspace code with minimum subspace distance $d$. Moreover, let $\Delta_E$ and $\Delta_F$ be the subspace weights of distinct $E,F \in \mathcal C$, respectively, such that  $\Delta_E \le w_S(A)$ for all $A \in \mathcal C \backslash \{F\}$ and $\Delta_F \le w_S(A)$ for all $A \in \mathcal C \backslash \{E\}$. If $\dim(A \cap B)=0$ for all $A,B \in \mathcal C$ with $A \ne B$ then $d=\Delta_E + \Delta_F$.
\end{theorem}

\begin{corollary}
Let $\mathcal C$ be an $(n,M,d,k)_q$ Grassmannian code. If $\dim(A \cap B)=0$ for all $A,B \in \mathcal C$ with $A \ne B$ then $d=2k$.
\end{corollary}

\begin{definition}
A function $w:\mathcal P_q(n) \rightarrow \RR$ is egalitarian if
	\begin{enumerate}
    	\item[(E)] there exists a $\Gamma \in \RR$ such that for any nonempty subset $\mathcal U$ of $\mathcal P_q(n)$,  \[\displaystyle \sum_{S\in \mathcal U}{w(S)}=\Gamma |U|.\]
	\end{enumerate}
\end{definition}

\begin{theorem}
The subspace weight is not egalitarian.
\label{subspace_weight}
\end{theorem}

\begin{IEEEproof} Consider two subsets $A$ and $B$ of $\mathcal P_q(n)$ with different dimensions. Suppose $\dim A = k_1$ and $\dim B = k_2$.
If $\dim A = k_1$, by definition, $\Gamma = k_1$.
If $\dim B = k_2$, $\Gamma = k_2$.
Hence the subspace weight is not egalitarian.
\end{IEEEproof}

\begin{remark}
Theorem \ref{subspace_weight} states that, in general, the subspace weight is not egalitarian. However, it is clear that, in the Grassmannian $\mathcal G _q (n,k)$, the average value $\Gamma$ is equal to $k$. Hence in the Grassmannian, the subspace weight is egalitarian.
\end{remark}
\section{Summary and Conclusion}
\label{sec:sum}
In this research we highlighted the role of one-sided ideals of the non-commutative matrix ring $M_2(\FF_q)$ as linear matrix codes in the construction of subspace codes, specifically Grassmannian codes, whose parameters are completely determined by the ideal. 

Weight properties of rank-metric codes and subspace codes were subsequently examined. The rank weight is not egalitarian nor homogeneous. Similarly the egalitarian property was defined on subspace codes. It turned out that the subspace weight is not egalitarian in general, but it is egalitarian in the Grassmannian.
\section{Acknowledgement}
The authors would like to thank Dixie F. Falcunit, Jr. for helpful discussions.

\end{document}